\documentclass{hep99}

\usepackage{cite}
\usepackage{epsfig}

\def\beq{\begin{equation}}
\def\eeq{\end{equation}}
\def\beqar{\begin{eqnarray}}
\def\eeqar{\end{eqnarray}}
\def\barr#1{\begin{array}{#1}}
\def\earr{\end{array}}
\def\bfi{\begin{figure}}
\def\efi{\end{figure}}
\def\btab{\begin{table}}
\def\etab{\end{table}}
\def\bce{\begin{center}}
\def\ece{\end{center}}

\def\text{\textstyle}

\def\si{\sigma}

\def\De{\Delta}

\def\reffi#1{\mbox{Fig.~\ref{#1}}}

\def\citere#1{\mbox{Ref.~\cite{#1}}}

\def\mathswitchr#1{\relax\ifmmode{\mathrm{#1}}\else$\mathrm{#1}$\fi}

\newcommand{\PW}{\mathswitchr W}

\newcommand{\PA}{\mathswitchr A}

\newcommand{\Ph}{\mathswitchr h}

\newcommand{\Pt}{\mathswitchr t}

\def\mathswitch#1{\relax\ifmmode#1\else$#1$\fi}

\newcommand{\MW}{\mathswitch {M_\PW}}

\newcommand{\Mt}{\mathswitch {m_\Pt}}

\newcommand{\mh}{\mathswitch {m_\Ph}}

\newcommand{\MA}{\mathswitch {M_\PA}}

\newcommand{\sweff}{\sin^2 \theta_{\mathrm{eff}}}

\def\tb{\tan\beta}

\newcommand{\mste}{m_{\tilde{\Pt}_1}}
\newcommand{\mstz}{m_{\tilde{\Pt}_2}}

\newcommand{\mt}{\Mt}

\newcommand{\Stop}{\tilde{t}}

\newcommand{\tst}{\theta_{\tilde{\Pt}}}

\newcommand{\tsf}{\theta\kern-.20em_{\tilde{f}}}
\newcommand{\tsfp}{\theta\kern-.20em_{\tilde{f}\prime}}
\newcommand{\tsq}{\theta\kern-.15em_{\tilde{q}}}
\newcommand{\lsim}
{\;\raisebox{-.3em}{$\stackrel{\displaystyle <}{\sim}$}\;}

\newcommand{\cp}{{\cal CP}}

\newcommand{\VL}{\left( \begin{array}{c}}
\newcommand{\VR}{\end{array} \right)}
\newcommand{\ML}{\left( \begin{array}{cc}}
\newcommand{\MLd}{\left( \begin{array}{ccc}}
\newcommand{\MLv}{\left( \begin{array}{cccc}}
\newcommand{\MR}{\end{array} \right)}

\newcommand{\gev}{\,\, \mathrm{GeV}}

\newcommand{\BC}{\begin{center}}
\newcommand{\EC}{\end{center}}
\newcommand{\BE}{\begin{equation}}
\newcommand{\EE}{\end{equation}}
\newcommand{\BEA}{\begin{eqnarray}}
\newcommand{\BEAnn}{\begin{eqnarray*}}
\newcommand{\EEA}{\end{eqnarray}}
\newcommand{\EEAnn}{\end{eqnarray*}}

\newcommand{\id}{{\rm 1\kern-.12em
\rule{0.3pt}{1.5ex}\raisebox{0.0ex}{\rule{0.1em}{0.3pt}}}}

\def\draftdate{\relax}
\def\mda{\relax}
\def\mua{\relax}
\def\mla{\relax}
\def\draft{
\def\thtystars{******************************}
\def\sixtystars{\thtystars\thtystars}
\typeout{}
\typeout{\sixtystars**}
\typeout{* Draft mode!
         For final version remove \protect\draft\space in source file
*}
\typeout{\sixtystars**}
\typeout{}
\def\draftdate{\today}
\def\mua{\marginpar[\boldmath\hfil$\uparrow$]%
                   {\boldmath$\uparrow$\hfil}%
                    \typeout{marginpar: $\uparrow$}\ignorespaces}
\def\mda{\marginpar[\boldmath\hfil$\downarrow$]%
                   {\boldmath$\downarrow$\hfil}%
                    \typeout{marginpar: $\downarrow$}\ignorespaces}
\def\mla{\marginpar[\boldmath\hfil$\rightarrow$]%
                   {\boldmath$\leftarrow $\hfil}%
                    \typeout{marginpar:
$\leftrightarrow$}\ignorespaces}
\def\Mua{\marginpar[\boldmath\hfil$\Uparrow$]%
                   {\boldmath$\Uparrow$\hfil}%
                    \typeout{marginpar: $\Uparrow$}\ignorespaces}
\def\Mda{\marginpar[\boldmath\hfil$\Downarrow$]%
                   {\boldmath$\Downarrow$\hfil}%
                    \typeout{marginpar: $\Downarrow$}\ignorespaces}
\def\Mla{\marginpar[\boldmath\hfil$\Rightarrow$]%
                   {\boldmath$\Leftarrow $\hfil}%
                    \typeout{marginpar:
$\Leftrightarrow$}\ignorespaces}
\overfullrule 5pt
\oddsidemargin -15mm
\marginparwidth 29mm
}

\begin{document}

\onecolumn

\thispagestyle{empty}

\def\thefootnote{\fnsymbol{footnote}}

\null
\hfill CERN--TH/2000--006\\
\null
\hfill hep-ph/0001044
\vskip 1.8cm
\begin{center}
{\Large \bf The MSSM at present and future colliders
\footnote{Presented at the ``International Europhysics
  Conference on High Energy Physics'', Tampere, July 1999.}
}
\vskip 3.5em
{\large
{\sc Georg Weiglein
}\\[3ex]
{\normalsize \it CERN, TH Division, CH--1211 Geneva 23, Switzerland}
}
\vskip 2em
\end{center} \par
\vskip 1.6cm
%\vfil
\begin{center}
{\bf Abstract} \\[.8em]%\par
\end{center}
A brief overview over the phenomenology of the MSSM at present and future
colliders is given. The complementarity of indirect tests of the model
via precision observables and of the information from the direct production
of SUSY particles is emphasized. If the lightest Higgs boson of the MSSM
will be detected, its mass will also play an important role as a
precision observable.
\par
\vskip 1cm
\null

\setcounter{page}{0}

\clearpage

\twocolumn

%%%%%%%%%%%%%%%%%%%%%%%%%%%%%%%%%%%%%%%%%%%%%%%%%%%%%%%%%%%%%%%%%%%
%%%%%%%%%%%%%%%%%%%%%%%%%%%%%%%%%%%%%%%%%%%%%%%%%%%%%%%%%%%%%%%%%%%

\title{The MSSM at present and future colliders}

\author{Georg Weiglein}
\address{CERN, TH Division, CH--1211 Geneva 23, Switzerland\\[3pt]
E-mail: {\tt Georg.Weiglein@cern.ch}}

\abstract{
A brief overview over the phenomenology of the MSSM at present and future
colliders is given. The complementarity of indirect tests of the model
via precision observables and of the information from the direct production
of SUSY particles is emphasized. If the lightest Higgs boson of the MSSM
will be detected, its mass will also play an important role as a
precision observable.
}

\maketitle

% Text of footnotes comes after \maketitle
%\fntext{1}{E-mail: tony.cox@ioppublishing.co.uk}
%\fntext{2}{E-mail: jim.revill@ioppublishing.co.uk}
%\fntext{\dag}{Here is a footnote.}

\section{Introduction}

Supersymmetric (SUSY) theories possess very appealing theoretical properties
(for a review, see e.g.\ \citere{mssm})
and can certainly be called the currently best motivated extensions of
the Standard Model (SM). Their minimal realization, the Minimal 
Supersymmetric Standard Model (MSSM), postulates superpartners to the 
SM fields and requires an enlarged Higgs sector with two Higgs doublets
giving rise to five physical Higgs-boson states. 

While the MSSM is
minimal in the sense of its particle content, in its unconstrained form
(i.e.\ without specific assumptions about the SUSY-breaking mechanism) it 
introduces more than 100 free parameters (masses, mixing angles, etc.)
in addition to the SM parameters.
If low-energy Supersymmetry turns out to be realized in nature and 
superpartners will be found at the present or the next generation of
colliders, the determination of the MSSM parameters will be a 
very demanding task, both from the experimental and the theoretical
side. A precise determination of the model parameters will not only be 
important in order to investigate whether the MSSM is consistent with the 
data, but also to infer possible patterns of the underlying SUSY-breaking
mechanism from the spectrum of the SUSY particles.

In this context it will be important to take advantage of all possible 
sources of information, i.e.\ both from the direct production of SUSY
particles and from indirect constraints on the model via precision
observables.

%%%%%%%%%%%%%%%%%%%%%%%%%%%%%%%%%%%%%%%%%%%%%%%%%%%%%%%%%%%%%%
%%%%%%%%%%%%%%%%%%%%%%%%%%%%%%%%%%%%%%%%%%%%%%%%%%%%%%%%%%%%%%

%\section{Production and decay processes for SUSY particles}
\section{Direct production of SUSY particles}

A detailed investigation of the production and decay processes of SUSY 
particles is indispensable for the SUSY searches at present and future
colliders, as the main background for SUSY signals will often be
SUSY itself.
%It is complicated by the fact that the main background for
%SUSY signals is often SUSY itself and thus analyses within the
%unconstrained MSSM can be very involved.
%
For production processes at hadron colliders QCD corrections are very
important. In general they give rise to a considerable enhancement of the
production cross sections (for recent reviews, see \citere{hadprod}).
Complementary to the hadron colliders Tevatron and LHC, where in
particular the latter has a large discovery potential for a wide range
of SUSY processes, an $e^+ e^-$ linear collider provides high-precision
information for all kinematically accessible SUSY particles (see
\citere{lcrep} for an overview). In the context of constraining the
parameters of the model it can be very useful to take advantage of
polarization of the $e^-$ and also the $e^+$ beam or to
study spin correlations in the production and subsequent decay of SUSY
particles (see \citere{polspincorr} and references therein).

%%%%%%%%%%%%%%%%%%%%%% FIGURE %%%%%%%%%%%%%%%%%%%%%%%%%%%%%%%%
\begin{figure}
\begin{center}
\mbox{
\epsfig{figure=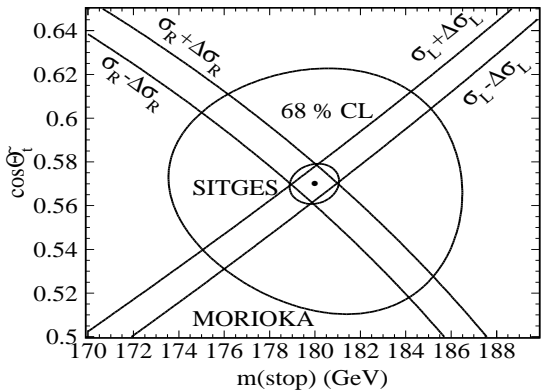,width=7.5cm,height=5.9cm}}
\end{center}
\caption[]{Parameter determination in the scalar top sector of the MSSM
at a linear collider with $\sqrt s =500$~GeV and a polarization of 
$+0.9$ (cross section $\si_{\mathrm{R}}$) and $-0.9$ (cross section 
$\si_{\mathrm{L}}$) of the $e^-$ beam. For the simulation 
an integrated luminosity of ${\cal L}=500$~fb$^{-1}$ (small contour)
and of ${\cal L}=10$~fb$^{-1}$ (large contour) has been assumed (taken
from \citere{sop}).
}
\label{fig:stopdir}
\end{figure}
%\caption{\label{fig:ellipse} Error bands and the corresponding error
%ellipse
%as a function of \mt\ and \cost\ for $\sqrt s =500$~GeV
%and ${\cal L}=500$~fb$^{-1}$.
%The dot corresponds to $\mt=180$~GeV and $\cost=0.57$.
%%The resulting error ellipse is also shown.
%The errors from the 10~fb$^{-1}$ Morioka study were
%$\Delta\mt$\,$=$\,7~GeV and $\Delta\cost$\,$=$\,0.06,
%while the 500~fb$^{-1}$ Sitges study
%gives $\Delta\mt=1$~GeV and $\Delta\cost=0.009$.}

%%%%%%%%%%%%%%%%%%%%%% FIGURE %%%%%%%%%%%%%%%%%%%%%%%%%%%%%%%%

As an example for the production of scalar top quarks at a linear
collider with $\sqrt s =500$~GeV~\cite{stopsec,sop}, \reffi{fig:stopdir} 
shows the determination of the mass of the lightest scalar top quark 
and the mixing angle in the $\Stop$~sector from the cross sections with 
polarization of the $e^-$ beam of $\pm 0.9$.
As shown in the figure, for 
an integrated luminosity of ${\cal L}=500$~fb$^{-1}$ a very precise
measurement could be possible.

%%%%%%%%%%%%%%%%%%%%%%%%%%%%%%%%%%%%%%%%%%%%%%%%%%%%%%%%%%%%%%
%%%%%%%%%%%%%%%%%%%%%%%%%%%%%%%%%%%%%%%%%%%%%%%%%%%%%%%%%%%%%%

\section{The lightest Higgs boson in the MSSM}

In contrast to the SM, the mass of the lightest $\cp$-even Higgs boson in 
the MSSM, $\mh$, is not a free parameter, but is calculable from the other
parameters of the model. It is bounded to be smaller than
the Z-boson mass at the tree level. This bound, however, is strongly 
affected by large radiative corrections. The dominant 
%one-loop corrections~\cite{mhiggs1} arise from the top and scalar top 
%sector of the MSSM via terms of the form $\GF \mt^4 \ln(\mste \mstz/\mt^2)$. 
corrections arise from the %top and scalar top sector of the
$t$--$\Stop$~sector of the MSSM.
%MSSM~\cite{mhiggs1}. 
At two-loop order an upper bound on $\mh$ of about 
$\mh \lsim 135$~GeV is obtained~\cite{mhiggs2}. 

This upper bound on $\mh$ is a definite and robust prediction of the
MSSM, which can be tested at the present and the next generation
of colliders. By comparing the present experimental limit on $\mh$ from
the search at LEP2 with the theoretical result for the upper bound on $\mh$ 
in the MSSM as a function of $\tb$, it is possible to derive constraints
on $\tb$. For recent analyses in this context, see \citere{tbanal1,tbanal2}.

If the lightest $\cp$-even Higgs boson of the MSSM will be found,
%at the present or the next generation of colliders, 
its mass will be determined with high
precision. The prospective accuracy at the LHC is 
$\De\mh = 0.2$~GeV~\cite{lhcmh}, while at a future linear collider an
accuracy of even $\De\mh = 0.05$~GeV~\cite{lcmh} could be achievable.
%This level of experimental precision will allow very sensitive tests of
%the model.

%%%%%%%%%%%%%%%%%%%%%%%%%%%%%%%%%%%%%%%%%%%%%%%%%%%%%%%%%%%%%%%%%%%%%%%%
%%%%%%%%%%%%%%%%%%%%%%%%%%%%%%%%%%%%%%%%%%%%%%%%%%%%%%%%%%%%%%%%%%%%%%%%

\section{Precision tests of the MSSM}

Complementary to the direct production processes of SUSY particles, 
constraints on the model can also be obtained from the 
virtual contributions of SUSY particles to SM processes. In global fits
to the electroweak data taken at LEP, SLC and the Tevatron the fit
quality in the MSSM is similar to the SM case~\cite{fits}. %Furthermore,
In the low energy regime rare $B$ decays have turned out to be 
sensitive probes for physics beyond the SM~\cite{bsg}. 
%An example of 
%a SM process at a future linear collider, where virtual SUSY 
%contributions can have a sizable effect, is $e^+e^- \to t \bar t$. 
%If the scalar top quarks are sufficiently light, electroweak SUSY loop 
%corrections can reach up to 10\% in this case~\cite{schapp}.

\begin{figure}
\begin{center}
\epsfig{figure=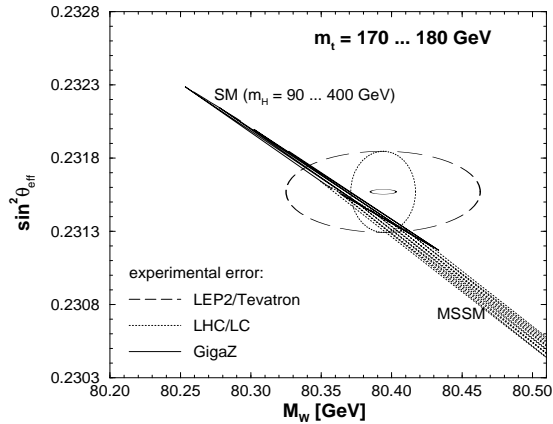,height=5.7cm,width=7.2cm}
\end{center}
\caption{Theoretical prediction of the SM and the MSSM in the
$\sweff$--$\MW$-plane compared with expected experimental accuracies at
LEP2/Tevatron, the LHC and GigaZ.
\label{fig:swmw}}
\end{figure}

Of particular importance for deriving indirect constraints on the MSSM 
are the precision observables $\MW$, $\sweff$, and in the future
possibly also $\mh$. %, as discussed above. 
In \reffi{fig:swmw} the SM and the MSSM predictions
for $\MW$ and $\sweff$, based on the complete one-loop results and the
leading higher-order QCD and electroweak corrections (see
\citere{smsusycorr} and references therein), are
%corrections~\cite{delr,mssm1l,sm2lqcd,sm3lqcd,sm2lmt4,mssm2lqcd}, are 
compared with the experimental accuracy obtainable at LEP2, SLC and the
Tevatron as well as with prospective future accuracies at the LHC and
at a high-luminosity linear collider in a dedicated low-energy run
(GigaZ)~\cite{sitgesPO}. 
The experimental accuracies assumed in \reffi{fig:swmw} for
LEP2/Tevatron, LHC and GigaZ are $\De \MW = 30$~MeV, 15~MeV, 6~MeV and
$\De \sweff = 1.8 \times 10^{-4}$, $1.8 \times 10^{-4}$, $1 \times 10^{-5}$, 
respectively.
%(see \citere{sitgesPO} and references therein). 
The allowed region of the SM prediction corresponds to varying $\mh$ in 
the interval 
$90 \gev \leq \mh \leq 400 \gev$ and $\mt$ within its present experimental 
uncertainty, while in the region of the MSSM prediction besides the 
uncertainty of $\mt$ also the SUSY parameters are varied. As can be seen
in the figure, the precision observables $\MW$ and $\sweff$ provide a
very sensitive test of the theory, in particular in the case of the 
GigaZ accuracy. It should be noted that with a future detection of the
Higgs boson, a prospective reduction on the experimental error of 
$\mt$ to $\De \mt = 2$~GeV at the LHC~\cite{lhcmh} and 
$\De \mt = 0.2$~GeV at a linear collider~\cite{lcmh}, 
and with the possible detection of SUSY particles the 
allowed range of the theory prediction in \reffi{fig:swmw} will be
drastically reduced. 
%The precision observables will thus be a very
%stringent consistency test of the theory with a high sensitivity to
%possible effects of new physics.

%%%%%%%%%%%%%%%%%%%%%% FIGURE %%%%%%%%%%%%%%%%%%%%%%%%%%%%%%%%
%%%%%%%%%%%%%%%%%%%%%% FIGURE %%%%%%%%%%%%%%%%%%%%%%%%%%%%%%%%

%If the lightest $\cp$-even Higgs boson of the MSSM will be found, its
%mass will be a further precision observable suitable for testing the
%MSSM. 
The prediction for $\mh$ within the MSSM is particularly sensitive to the
parameters in the $t$--$\Stop$~sector, while in the region
of large $\MA$ and large $\tan\beta$ (giving rise to Higgs masses beyond
the reach of LEP2) the dependence on the latter two parameters is
relatively mild. A precise measurement of $\mh$ can thus be used to 
constrain the parameters in the %scalar top sector of the MSSM. 
$t$--$\Stop$~sector of the MSSM.

%%%%%%%%%%%%%%%%%%%%%% FIGURE %%%%%%%%%%%%%%%%%%%%%%%%%%%%%%%%
\begin{figure}
\begin{center}
\epsfig{figure=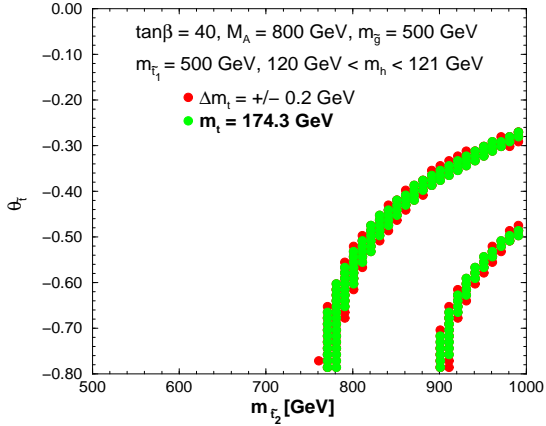,height=5.7cm,width=7.2cm}
\end{center}
\caption{Indirect constraints on the parameters of the scalar top sector
of the MSSM from a precision measurement of $\mh$.
\label{fig:mhconstr}}
\end{figure}
%%%%%%%%%%%%%%%%%%%%%% FIGURE %%%%%%%%%%%%%%%%%%%%%%%%%%%%%%%%

In \reffi{fig:mhconstr} it is assumed that the mass of the lightest
scalar top quark, $\mste$, is known with high precision, while the mass
of the heavier scalar top quark, $\mstz$, and the mixing angle $\tst$
are treated as free parameters. The Higgs boson mass is assumed to be
known with an experimental precision of $\pm 0.5$~GeV (and a
hypothetical value for the central value of $\mh$ is considered), 
and $\De \mt = 0.2$~GeV is used. The figure shows that the values of 
$\mstz$, $\tst$ which are compatible with a Higgs-mass prediction of 
$\mh = 120.5 \pm 0.5$~GeV are given by two narrow bands in the 
$\mstz$--$\tst$~plane (the bands corresponding to smaller and larger
values of $\mstz$ are related to smaller and larger values of the
off-diagonal entry in the scalar top mixing matrix, respectively). 
The uncertainty of $\De \mt = 0.2$~GeV assumed in \reffi{fig:mhconstr}
is seen to have only a marginal effect. Combining the
constraints on the parameters in the scalar top sector with the
constraints from the precision observables $\MW$ and $\sweff$ and with 
the information from the direct production of the scalar top quarks
(see \reffi{fig:stopdir}) will clearly lead to a very sensitive test
of the MSSM. In \reffi{fig:mhconstr} the theoretical uncertainty in the
Higgs-mass prediction from unknown higher-order contributions and the 
parametric uncertainty related to all parameters besides $\mstz$,
$\tst$, and $\mt$ has been neglected. In a more realistic analysis these
uncertainties, in particular
%will have to be taken into account and in particular the 
%uncertainty from 
the dependence on the other SUSY parameters according
to the available experimental information on these parameters, will have
to be taken into account.\\[.3em] %included.

The author thanks S.~Heinemeyer, W.~Hollik and P.M.~Zerwas for
collaboration and W.~de Boer, J.~Erler, M.~Kr\"amer, S.~Kraml,
G.~Moortgat-Pick, C.~Schappacher, A.~Sopczak and U.~Schwickerath for 
useful communications.

%%%%%%%%%%%%%%%%%%%%%%%%%%%%%%%%%%%%%%%%%%%%%%%%%%%%%%%%%%%%%%%%%%%%%%%%
%%%%%%%%%%%%%%%%%%%%%%%%%%%%%%%%%%%%%%%%%%%%%%%%%%%%%%%%%%%%%%%%%%%%%%%%

\end{document}